# Frequency-tunable nano-oscillator based on Ovonic Threshold Switch (OTS)

Seon Jeong Kim[1,2], Seong Won Cho[1,3], Hyejin Lee[1,4], Jaesang Lee[1,3], Tae Yeon Seong[2], Inho Kim[1], Jong-Keuk Park[1], Joon Young Kwak[1], Jaewook Kim[1], Jongkil Park[1], YeonJoo Jeong[1], Gyu Weon Hwang[1], Kyeong Seok Lee[1], Suyoun Lee[1,5*]

[1]*Center for Neuromorphic Engineering, Korea Institute of Science and Technology, Seoul 02792, Korea*

[2]*Department of Material Science and Engineering, Korea University, Seoul 02841, Republic of Korea*

[3]*Korea Department of Materials Science and Engineering, Seoul National University, Seoul 08826, Korea*

[4]*Institute of Physics and Applied Physics, Yonsei University, Seoul 03722, Korea*

[5]*Division of Nano & Information Technology, Korea University of Science and Technology, Daejon 34316, Korea*

**Nano-oscillator devices are gaining more and more attention as a prerequisite for developing novel energy-efficient computing systems based on coupled oscillators. Here, we introduce a highly scalable, frequency-tunable nano-oscillator consisting of one Ovonic threshold switch (OTS) and a field-effect transistor (FET). It is presented that the proposed device shows an oscillating behavior with a natural frequency ($f_{nat}$) adjustable from 0.5 to 2 MHz depending on the gate voltage applied to the FET. In addition, under a small periodic input, it is observed that the oscillating frequency ($f_{osc}$) of the device is locked to the frequency ($f_{in}$) of the input when $f_{in} \approx f_{nat}$, demonstrating the so-called "synchronization" phenomenon. It also shows the phase lock of the combined oscillator network using circuit simulation, where the phase relation between the oscillators can be controlled by the coupling strength. These results imply that the proposed device is promising for applications in oscillator-based computing systems.**

## Introduction

Electronic oscillators are commonly found in everyday circuits ranging from communication devices to conventional computing devices. Interest in oscillators is increasing in recent years as they provide building blocks for the development of non-traditional and energy-efficient computing systems based on coupled oscillators[1] such as Ising machine (IM)[2,3] and oscillator-based reservoir computing[4,5]. Conventional oscillators based on digital inverter circuits or coupled inductor-capacitor circuits are problematic for applications in these oscillator-based computing systems, as they usually require too large areas to build large-scale computing systems. For this reason, much attention has been paid to novel nanoscale oscillator devices such as spin-torque nano-oscillators (STNOs)[6-9], spin-Hall nano-oscillators (SHNOs)[10-12], and relaxation oscillators (ROs)[13-15]. Nevertheless, these devices are considered questionable because of some drawbacks: For STNO or SHNO, it not only needs a magnetic tunnel junction (MTJ) consisting of a rather complex stack but also needs an external magnetic field for the operation resulting in an issue of the energy consumption. For conventional RO, it requires a capacitor usually allocating a large area, which reduces scalability.

Here, we propose a new nano-oscillator as a solution to these problems. Fig. 1a shows the circuit diagram of a nano-oscillator suggested in this work, which consists of one Ovonic Threshold Switch (OTS) and one resistor(1OTS+1R). An OTS device consisting of an amorphous chalcogenide sandwiched between the electrodes shows reversible electrical switching[16], where the switching to the ON(OFF) state is considered to arise from charging(discharging) of the trap states inside the amorphous chalcogenide[17-25]. Two characteristic voltages are the threshold voltage ($V_{th}$) and the holding voltage ($V_H < V_{th}$, see the inset in Fig. 1a), at which the resistance of the OTS turns low ($R_{on}$) and high ($R_{off}$) with a very high on/off ratio (~$10^6$), respectively[26-29]. When a voltage bias ($V_{bias} > V_{th}$) is applied to the series connection of OTS and load resistor ($R_L$), if $R_L$ is set such that $R_{on} \ll R_L \ll R_{off}$, it is easy to predict the oscillating behavior of the device considering voltage-dividing between the OTS and the load resistor.

The duration of the ON state is determined by the discharge time of the trap states inside the amorphous chalcogenide, which is dependent on $R_L$. If $R_L$ is replaced by a FET(field effect transistor, 1OTS+1FET) whose resistance can be modulated by the gate voltage ($V_G$), the oscillation frequency can be adjusted. This principle of operation differs somewhat from that of conventional ROs in that the proposed 1OTS+1R device does not require a capacitor, which gives it much-improved scalability.

One of the important properties of oscillators when applying oscillators and coupled oscillator networks in computing devices is the "synchronization" function[30]. That is, for an oscillator that receives an external periodic force with a frequency of $f_{ext}$, the frequency of the oscillator ($f_{osc}$) is modified to the frequency of the external force within a certain range of $f_{ext}$. What's more, when synchronizing, the phase of the oscillator is slipped and locked to the phase of the external force. These frequency- and phase-locking properties of the coupled oscillator network constitute the basis of oscillator-based computing, which is considered promising for energy-efficient solvers of the NP-hard problems[31-38].

In this study, we have investigated the characteristics of the 1OTS+1FET oscillator, showing sustained oscillation under a DC bias, the possibility of frequency tuning through $V_G$ of the FET, and synchronization under a small AC modulation bias. We also used SPICE (Simulation Program with Integrated Circuit Emphasis) software to investigate the feasibility of a coupled-oscillator network of 1OTS+1FET oscillators for applications in oscillator-based computing systems. The following results show that the 1OTS+1FET oscillator is a promising candidate for the application in a large-scale computing system based on coupled oscillators.

## Results and Discussion

Fig. 1b shows a cross-sectional transmission electron microscope (TEM) image of an OTS device with $d$=300 nm. Fig. 1c shows the characteristic current-voltage (*I-V*) curves of the device, which is nominally the same as Fig. 1b. It shows the variation in repeated measurements (ten measurements) and shows the variation of $V_{th}$ within 2.5 ~ 3.0 V. On the other hand, it shows almost constant $V_H$ near 1 V with a negligible variation in repeated measurements.

To investigate the oscillating behavior of the proposed 1OTS+1R device, a DC voltage bias (=4 V) is applied to $V_{in}$ and the transient $V_{out}(t)$ is measured. Fig. 2a shows three representative $V_{out}(t)$ waveforms with $R_L$= 700 kΩ, 20 kΩ, and 0.7 kΩ from bottom to top, respectively. As expected, it clearly shows that the proposed 1OTS+1R device exhibits oscillating behavior over a certain range of $R_L$. For the analysis of the mechanism of the oscillating behavior, $V_{out}(t)$ is fitted to $V_{out}(t)=V_0+V_1\exp(-t/\tau_{dis})$ in each discharging period, where $V_0$ and $V_1$ are constants and $\tau_{dis}$ is the characteristic time for discharging. Assuming that the OTS device is a parallel connection of a resistor ($R_{ON}^{OTS}$) and a parasitic capacitor ($C_{para}$), $\tau_{dis}$ is given by $\tau_{dis} \cong R_L C_{para}$. The $R_L$-dependence of $\tau_{dis}$ has been measured as shown in Fig. 2b, from which $C_{para}$ is evaluated as ~236 pF from the slope in the $\tau_{dis}$ vs. $R_L$ curve. Given that the dielectric constant of GeSe is reported to be 13[26,39], the obtained $C_{para}$ is too large. In fact, by measuring the capacitance of OTS devices with various pore sizes, we confirm that the capacitance of the device is in the range of 2~4 pF depending on the pore size. These results imply that another mechanism is involved in the discharging process. It is known that the amorphous chalcogenide in the OTS device has a huge amount of traps (~$10^{21}$ /cm$^3$) and the ON state of the OTS is modeled as the state where electrons move above the trap while the trap states are filled[22,40]. After removing the bias, filled trap states must be emptied for the OTS to return to the OFF state. Therefore, we believe that the trap states in the OTS device play a similar role to the capacitor of a conventional relaxation oscillator. This property of the OTS device gives OTS-based oscillators improved scalability because a capacitor typically requires a large area.

To investigate the frequency-controllability of a 1OTS+1FET device, a DC voltage bias

(= 4 V) is applied to $V_{in}$ and the transient $V_{out}(t)$ is measured with varying $V_G$ applied to the gate of the FET. Fig. 3a shows $V_{out}(t)$ of a typical device with varying $V_G$, clearly showing that $f_{nat}$ of the oscillator can be controlled by $V_G$. That is, as $V_G$ increases, so does $f_{nat}$. The observed increase in $f_{nat}$ with $V_G$ is explained by the reduction in the aforementioned discharging period in the OTS device due to the decrease in resistance to ground. To quantify $f_{nat}$ at each $V_G$, the fast Fourier transform (FFT) is performed as shown in Fig. 3b. Here $f_{nat}$ is defined by the frequency of the primary peak while the satellite peaks are high order harmonics. Fig. 3c shows $f_{nat}$'s highly linear dependency on $V_G$. These results clearly show that the oscillation in the OTS-based oscillator is composed of fundamental-frequency components and its harmonics keeping other components negligible. In addition, it is observed that the primary peak in Fig. 3b is quite sharp and that $f_{nat}$ is adjustable in the range of 0.5 ~ 2 MHz, demonstrating the promising characteristics of OTS-based nano-oscillators.

In novel computing techniques based on coupled oscillators[1], an essential ingredient is the "synchronization". This means that the oscillator's frequency and phase are locked to each other as mentioned above. In a single oscillator device, it is demonstrated by the locking of the frequency and the phase to the external force which has a small oscillating component. To investigate the synchronization (sync.) characteristics, an external bias, a mixture of a DC (=4 V) and an AC (=0.23~0.27 V), is applied to the OTS-based nano-oscillator, and the output waveform is measured with changing the frequency ($f_{in}$) of the AC component. This has been repeated for different gate voltages and typical results obtained with $V_G$=-0.7 V are shown in Fig. 4a. Six representative waveforms are displayed; the bottom two (below synchronization), the middle two (in synchronization), and the top two (above synchronization). Note that the two waveforms in synchronization are clearly distinguishable in that they exhibit a very simple oscillating behavior. In contrast, the other four waveforms show a rather complex oscillating behavior with long-range oscillation envelopes. In addition, the FFT-amplitudes of these waveforms are shown in Fig. 4b. It shows the complex structure of the out-of-synchronization regime waveform, as opposed to the simple structure of the synchronized regime. In Fig. 4c, the measured oscillation frequency ($f_{osc}$, see Fig. 4b) is plotted as a function of $f_{in}$ at various gate voltages, showing that $f_{osc}$ coincides with $f_{in}$ in a certain range with its center and width depending on $V_G$. These results indicate that the frequency and the phase of our nano-oscillator are locked to an external bias, clearly showing synchronization.

We have also used SPICE software (LTspice, Linear Technology) to investigate the applicability of our oscillator devices in computing systems based on coupled oscillators. The OTS device is modeled as a parallel connection between a voltage-controlled switch and a capacitor. Using a network consisting of four identical oscillators (see Fig. 5a) and coupling resistors ($R_c$), we investigate the phase of each oscillator with varying $R_c$. All $R_c$s are assumed to have the same value for simplicity and the initial phase of each oscillator is chosen arbitrarily. In Fig. 5b~d, the waveforms of each oscillator are displayed when $R_c=5\times10^6$, 5000, and 50 Ω, respectively. In Fig. 5b, which shows the case of negligible coupling ($R_c$=5 MΩ), it can be seen that each oscillator behaves independently of each other. However, at the intermediate coupling strength ($R_c$=5 kΩ, Fig. 5c), note that the phase of oscillators becomes correlated after a while, showing that the adjacent devices oscillate with a phase difference of π (anti-phase oscillation). As the coupling becomes stronger ($R_c$=50 Ω, Fig. 5d), it turns out that all devices oscillate in approximately the same phase from the beginning (in-phase oscillation). With varying $R_c$ from 50 Ω to 5 kΩ gradually, it is observed that the phase difference between adjacent oscillators shows bistable at 0 and π, with a sharp transition around 200 Ω. These results mean that our devices are promising for applications in oscillator-based computing.

**Summary**

In this study, we have investigated the 1OTS+1FET composite device as a candidate nano-oscillator with improved scalability for applications in computing systems based on coupled oscillators. We have demonstrated that the proposed device shows robust oscillating behavior with a well-defined frequency, which can be controlled by the gate voltage applied to the FET. Analysis of the oscillation waveform shows that not only can the OTS device be modeled as a parallel connection of a voltage-controlled switch and a capacitor, but also that the capacitance originates from the trap states inside the amorphous chalcogenide, giving the OTS-based oscillators excellent scalability. In addition, the synchronization of the oscillator under the external force has been demonstrated, where the characteristics of the synchronization can be controlled by the FET. Finally, by simulating a simple coupled network composed of OTS-based oscillators, it has been shown that the phase of each oscillator in the network can be controlled by the coupling strength. In conclusion, these results show the promising characteristics of OTS-based oscillators in the application field of oscillator-based computing devices, and considering that OTS is a commercialized technology[27,41], we believe that OTS-based oscillators provide essential building blocks for developing large-scale oscillator-based computing systems.

## Methods

OTS devices are fabricated with a pore-type structure, where the pore size ($d$) is defined by the electron beam lithography in the range of 100~500 nm. $Ge_{37}Sn_{13}Se_{50}$ (GSS) with a thickness of 60 nm is used as a switching material because it is found to have a lower $V_{th}$ compared to $Ge_{50}Se_{50}$ (not shown here). It is deposited by co-sputtering technique (magnetron RF sputtering) using Ge, Sn, and $GeSe_2$ targets. The composition of GSS is calibrated using X-ray fluorescence (XRF) spectroscopy. Pt and TiN are used as the bottom and top electrodes, respectively.

To configure a frequency-adjustable nano-oscillator, an OTS device is connected in series to a FET device (LND150N3-G, Microchip Inc.). The characteristics of OTS devices and nano-oscillators have been investigated using an arbitrary function generator (AFG-3101, Tektronix), an oscilloscope (DPO-5104B, Tektronix), and a source-measure unit (2635B, Keithley). The capacitance of OTS devices is measured using an impedance analyzer (SI 1260, Solartron). We have tested several nano-oscillator devices using OTS devices of various pore sizes and found similar behavior. The results presented in this paper are mainly obtained by using OTS devices with $d$=300 nm except for the capacitance data, for which OTS devices with a larger pore size (2 ~ 40 μm).

## Acknowledgments

This work was supported by the Korea Institute of Science and Technology (KIST) through 2E30761 and by National Research Foundation program through NRF-2019M3F3A1A02072175.

**Author contributions**

S.L. designed and conceived the experiments with help from S.J.K., S.W.C., H.L., and J.L. S.J.K., S.W.C., H.L., and J.L. fabricated OTS and composite oscillator devices. S.J.K. and I.K. performed characterization of devices. S.L., J.-K.P., J.Y.K., J.K., and T.Y.S. participated in the analysis of the synchronization. J.P., Y.J., G.W.H., and K.S.L. performed a circuit analysis about the behavior of the oscillator device and a simulation of a coupled oscillator network. All authors discussed the data and participated in the writing of the manuscript.

**Competing interests**

The authors declare no competing interest.

# Figures

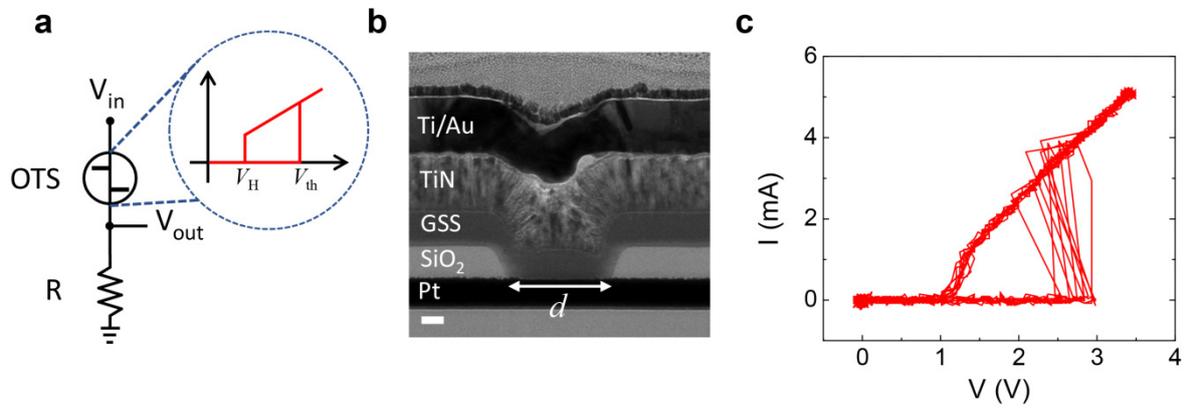

**Fig. 1 Frequency-tunable nano-oscillator based on OTS device. a** 1OTS+1R structure as a basic oscillator device. The inset shows a schematic *I-V* curve of an OTS device. **b** A cross-sectional TEM image of an OTS device, where GSS means $Ge_{37}Sn_{13}Se_{50}$ (scalebar=50 nm). **c** Characteristic *I-V* curve of an OTS device (ten repetitions).

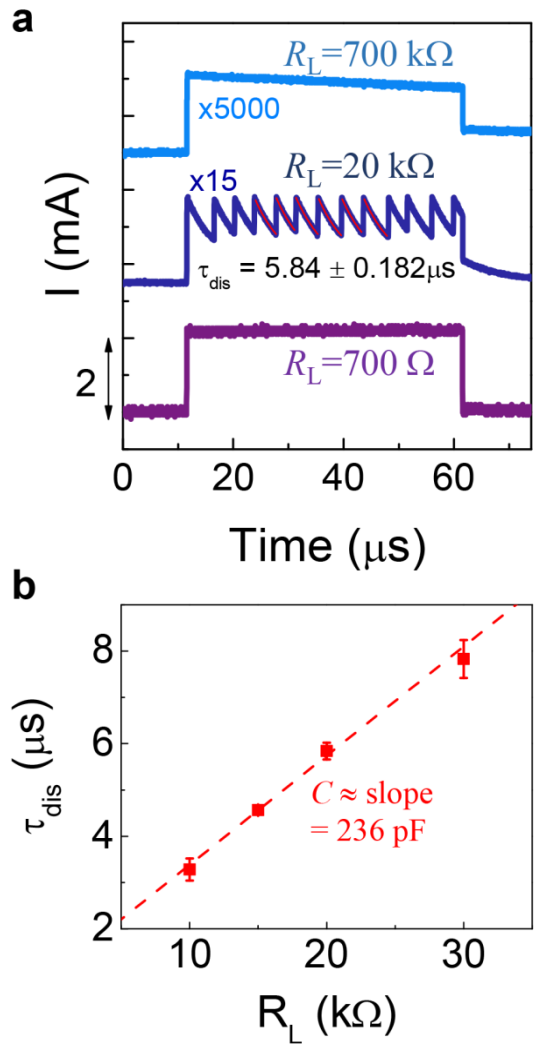

**Fig. 2. Basic oscillating behavior and analysis of the discharging period. a** Output waveforms (in current $I_{out}=V_{out}/R_L$) for three cases of $R_L$=700 kΩ, 20 kΩ, and 700 Ω from bottom to top, respectively. For the case of $R_L$=20 kΩ, the red curves are fitting curves to the data in discharging periods, where $I_{out}(t)=V_0+V_1\exp(-t/\tau_{dis})$ ($V_0$, $V_1$, and $\tau_{dis}$=constants). **b** $R_L$-dependence of $\tau_{dis}$. Errorbar represents the standard deviation obtained from 10 samples. The red solid line is a linear fit to the data, giving an estimation of the capacitance ($C$) at 236 pF.

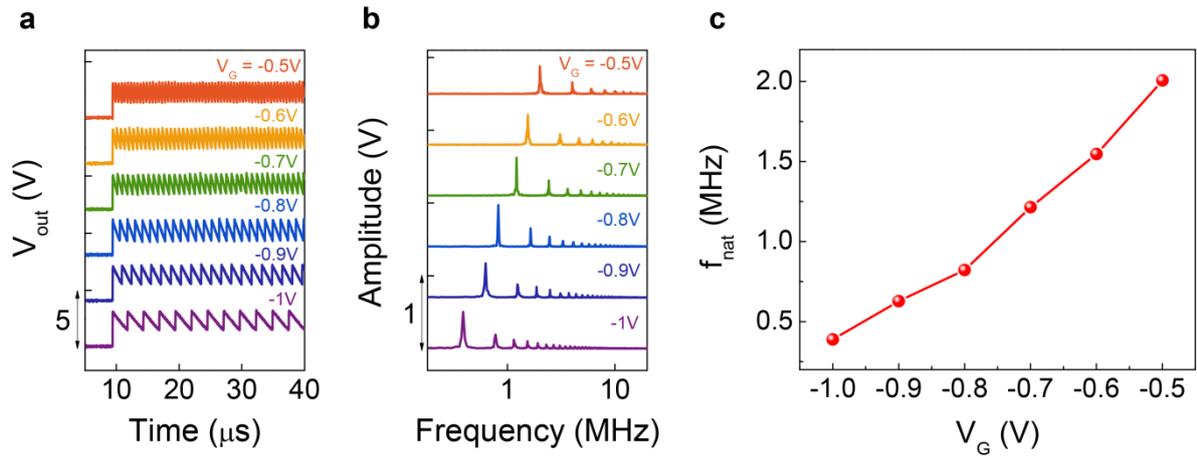

**Fig. 3. Frequency-tunability of the 1OTS+1FET oscillator. a** Output waveforms of the 1OTS+1FET oscillator for various $V_G$s. **b** Fast Fourier transform (FFT)-amplitude of the output waveforms. **c** Dependence of the natural frequency ($f_{nat}$) on $V_G$, where $f_{nat}$ is defined as the position of the primary peak in FFT.

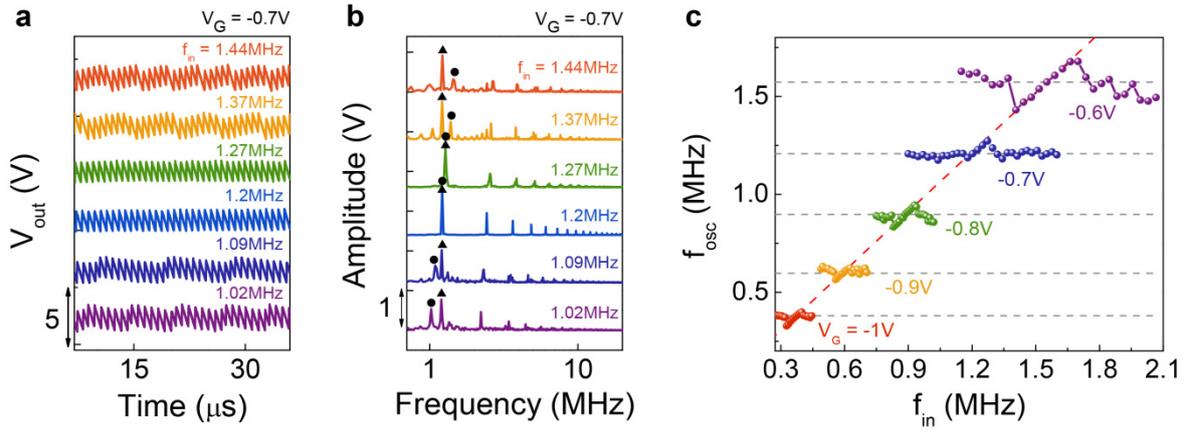

**Fig. 4. Synchronization of the 1OTS+1FET oscillator under a small oscillating input bias. a** Output waveforms of a representative 1OTS+1FET oscillator under an input bias $V_{bias}=V_{dc}+V_{ac}\sin(2\pi f_{in}t)$ for two $f_{in}$s below synchronization, another two $f_{in}$s in synchronization, and the rest two $f_{in}$s above synchronization from bottom to top, respectively. $V_G$ is set to -0.7 V. **b** FFT-amplitude of the waveforms, where circle and triangle represent the $f_{in}$ and $f_{osc}$, respectively. **c** Oscillation frequency ($f_{osc}$) as a function of $f_{in}$ for various $V_G$s.

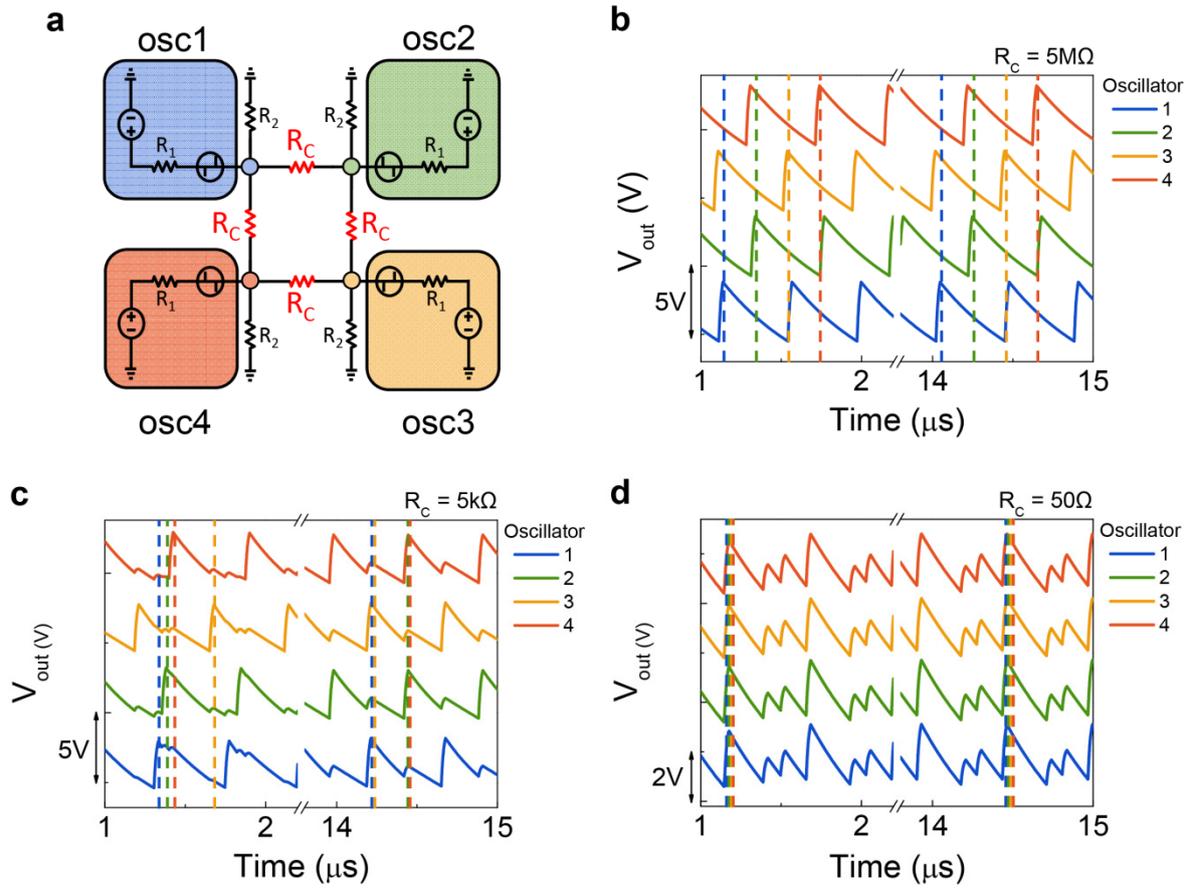

**Fig. 5. Spice simulation of dynamics of a network composed of four coupled oscillators.
a** A network of four coupled oscillators, where $R_1$ and $R_2$ are set to 1 kΩ and 5 kΩ, respectively. **b~d** Simulated waveforms of output voltages for $R_c$=5 MΩ, 5 kΩ, and 50 Ω, respectively. Vertical dashed lines represent positions of the peak of each waveform, which are drawn to help to see the change of the phase relation between oscillators with time.